# Concentric-electrode organic electrochemical transistors: case study for selective hydrazine sensing

S Pecqueur[*], S Lenfant, D Guérin, F Alibart and D Vuillaume
Institut d'Electronique, Micro-électronique et Nanotechnologie, CNRS, CS 60069, avenue Poincaré, 59652 Cedex, Villeneuve d'Ascq, France

* E-mail ; sebastien.pecqueur@*iemn.univ-lille1*.fr

**Abstract.** We report on hydrazine-sensing organic electrochemical transistors (OECTs) with a design consisting in concentric annular electrodes. The design engineering of these OECTs was motivated by the great potential of using OECT sensing arrays in fields such as bioelectronics. In this work, PEDOT:PSS-based OECTs have been studied as aqueous sensors, specifically sensitive to the lethal hydrazine molecule. These amperometric sensors have many relevant features for the development of hydrazine sensors, such as a sensitivity down to $10^{-5}$ M of hydrazine in water, an order of magnitude higher selectivity for hydrazine than for 9 other water soluble common analytes, the capability to recover entirely its base signal after water flushing and a very low voltage operation. The specificity for hydrazine to be sensed by our OECTs is caused by its catalytic oxidation at the gate electrode and enables increasing the output current modulation of the devices. This has permitted the device-geometry study of the whole series of 80 micrometric OECT devices with sub-20-nm PEDOT:PSS layers, channel lengths down to 1 $\mu$m and a specific device geometry of coplanar and concentric electrodes. The numerous geometries unravel new aspects of the OECT mechanisms governing the electrochemical sensing behaviours of the device, more particularly the effect of the contacts which are inherent at the micro-scale. By lowering the device cross-talking, micrometric gate-integrated radial OECTs shall contribute to the diminishing of the readout invasiveness and therefore promotes further the development of OECT biosensors.

## 1. Introduction

Polymeric-semiconductor devices have already shown great promises to substitute many technologies behind consumer products such as electronics [1], photovoltaics [2], lightings [3] and displays [4]. Recently, efforts have been devoted to employ the outstanding features of these plastic electronics technologies to create smarter sensors, such as electrolyte-gated organic field-effect transistors [5] or organic photodetectors [6]. More specifically, the rich chemistry of molecular and polymeric semiconductors offers the possibility to build biocompatible sensors. Indeed, organic electrochemical transistors are a type of electrolyte-gated transistors in which the constituting semiconducting ionomer blend has a good affinity with cations. As such, this platform is extensively studied as a bio-electronical interface to sense cellular action potentials [7-9]. Nowadays, one of the key challenges for this technological tool is to interpret complex readouts of dense networks. In order to control a whole array, one should first engineer the OECT units and second understand the mechanism of these units responsible for their sensitivity.

Efforts have been devoted to theoretically study on capacitive sensors having a radial symmetry [10] but these geometries are not extensively applied for practical applications. Here, we report results on OECTs with a circular geometry offering a higher sensing isotropy and a large geometry variation to interpret the device functioning. Also, the whole device comprises its individual gate electrode, which is locally patterned on the substrate in an overall diameter down to 100 $\mu$m. Despites the high current densities obtained at very low operated voltage, the dimension restrictions strongly limit the active areas of the sensor which hinders its gate-controlled response. Therefore, we activated the large-gate output-current modulation by the introduction of hydrazine. The OECTs show to be highly sensitive and selective to hydrazine. Hydrazine is a 0.1-ppm-level genotoxic for which the human carcinogenicity has not been evidenced but high caution is still recommended on inhalation-induced lung-cancer risks [11]. Thus, the ability to realise this kind of sensors is then of the clearest interest. Moreover, these OECTs have shown performances very depending on their geometry and we have explored in this study the hydrazine-enabled gate OECT on a population of 80 devices with different geometries to better understand their functioning.

**2. Experimental procedure**
The devices have been fabricated by e-beam lithography (see figure 1 a to d), adapted from a previously-reported process [12]. Platinum electrodes have been patterned by e-beam lithography and lift-off on top of a 200 nm thermally-grown $SiO_2$ / n-type Si substrate, using a resist of poly(methyl methacrylate). 10 nm of titanium were evaporated as an adhesion layer prior the 70 nm of platinum. Platinum has been chosen as a contact metal for both source/drain and gate because of its chemical inertia and its high work-function forming ohmic contacts with organic hole transport materials. Then, $SiO_2$ was functionalized with (3-glycidyloxypropyl)trimethoxysilane (22 mM-concentrated in a 40 mL 200:1-toluene/acetic-acid solution) as an adhesion promoter, prior the processing of a 2-$\mu$m-thick poly(chloro-*para*-xylylene) (or parylene C) capping layer by chemical vapor deposition using the Gorham process (Dichloro[2,2]-*para*-cyclophane was used as a precursor in a C20S deposition chamber from Comelec). Then, the device apertures were patterned by e-beam lithography with a sacrificial 3.7-$\mu$m-thick resist (a modified UV210 from Rohm and Haas Electronic Materials LLC) before dry-etching the surface with an oxygen plasma. Then, a surfactant (2 % v/v Micro-90 in deionized water) was dispensed on top of the substrate by spin-coating, before depositing a subsequent 2-$\mu$m-thick parylene C layer (under similar conditions as previously). Another e-beam lithography and dry-etching step, with similar conditions as previously, was carried out in order to etch the channel cavities and mask the gate apertures. The channel cavities were filled by a filtered and spincoated PEDOT:PSS solution (76 % v/v Clevios PH1000 Heraeus, 19 % v/v ethylene glycol, 4 % v/v dodecylbenzenesulfonic acid, 1 % v/v 3- glycidyloxypropyl)trimethoxysilane) and the underlying parylene C layer carefully stripped out with some adhesive tape, before an hour long hard baking at 140°C.

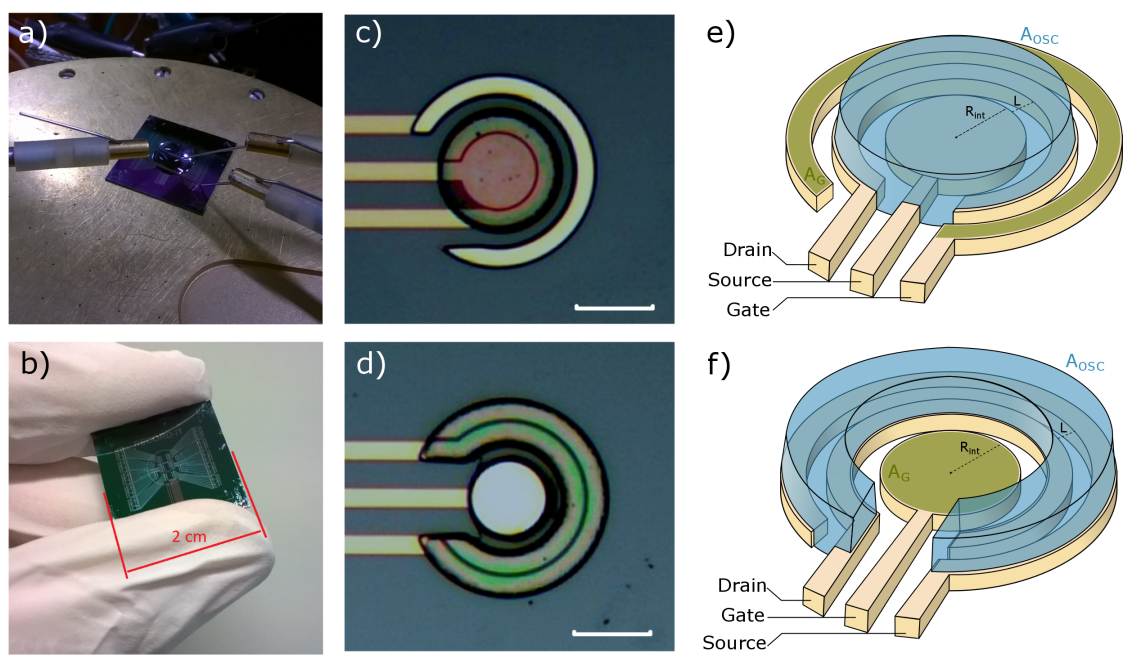

**Figure 1.** a) and b) Electrical setup including the 4-cm² OECT-patterned silicon substrate with the electrolyte drop dispensed on top of it. Source-centred (c and e) and the gate-centred (d and f) OECTs: Device microscope pictures (c and d, scale bar: 40 $\mu$m) and depiction of their architecture (e and f, the representation of the passivating parylene C has been omitted for sake of clarity).
The electrical characterization was performed on an Agilent 4155 parameter analyser, OECTs grounded at the gate. Before measurements and to ensure the highest reproducibility [13], the PEDOT:PSS patches were stressed 3 times with a grounded platinum wire gate at -700 mV in a 1 M KCl aqueous solution. Although the potentials are not referenced, we set the highest polarisation to 700 mV in absolute value, Pt gate grounded, in order to lower the risk of water electrolysis (no significant steady-state current density at the gate electrode was observed in the study).

## 3. Discussion on the OECT as a hydrazine sensor

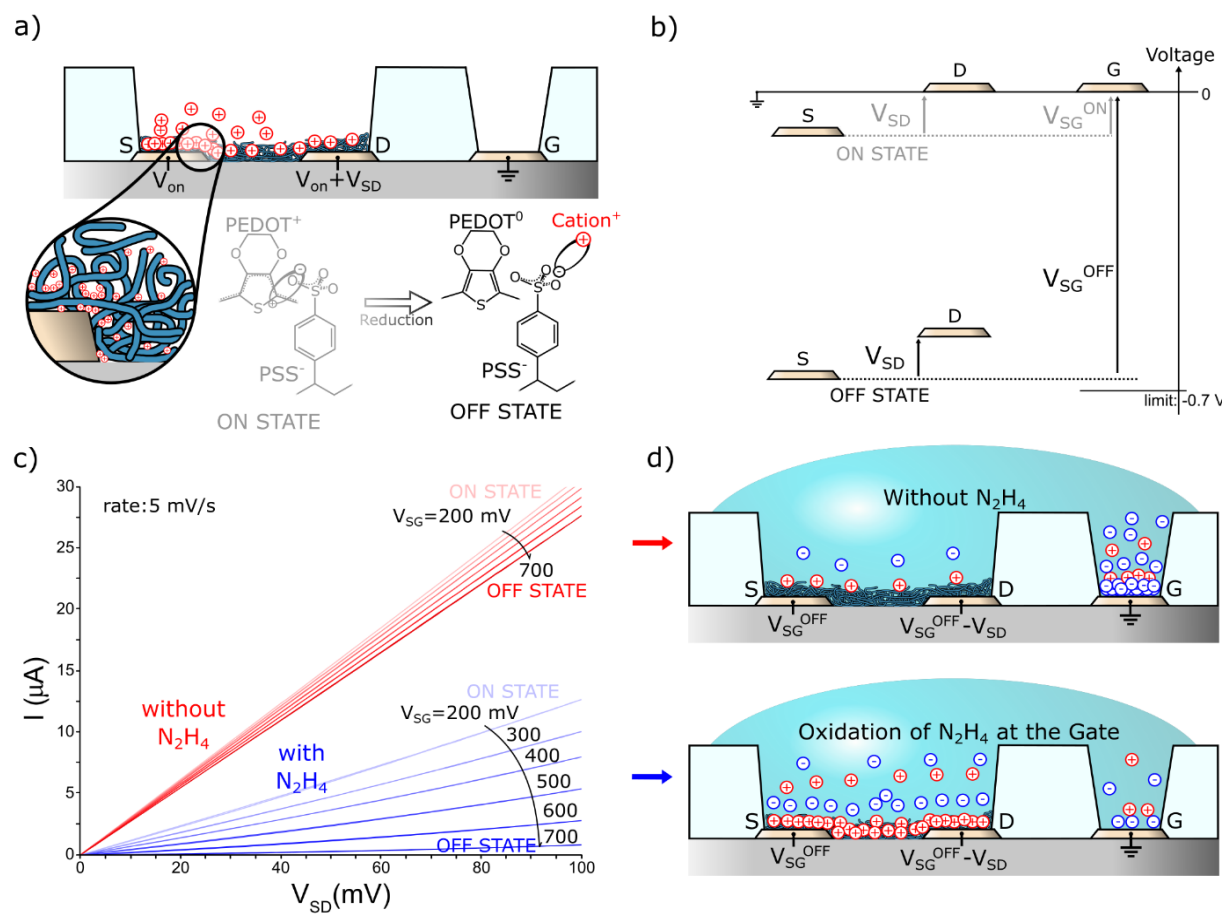

**Figure 2.** Behaviour of the OECT: a) Sketch of the widely accepted OECT operation model, implicating the participation of cations in the PEDOT dedoping mechanism upon channel/gate polarisation b) Diagram of the applied voltage biases, necessary in our experiments to read the OECT ON and OFF states c) Typical current-voltage characteristics of an OECT without and with hydrazine (conditions: 1 M KCl, 0.1 M $N_2H_4$, source-centred, L = 5 $\mu$m and $R_{int}$ = 20 $\mu$m) d) Schematic of two cases of different ionic distributions on an OECT under operation. At the top, the case for which the voltage drops mainly at the gate/electrolyte interface. At the bottom, the case for which the voltage drops mainly at the PEDOT/electrolyte interface.

Central symmetry OECTs have been fabricated on either a source-centred[1] or gate-centred configuration (see figure 1 e and f). The geometry of the devices was modulated by the channel length L (20, 10, 5, 2 and 1 $\mu$m) and the radius of the central electrode $R_{int}$ (50, 40, 30 and 20 $\mu$m).

---
[1] The term "source-centred" refers strictly to a device geometry and not the polarisation of the source with respect to the drain. This information is specified by the term "direct polarisation", opposed to "reversed polarisation".

Practically, the round-shaped PEDOT:PSS patches on the source-centred OECTs were patterned easier than the "C"-shaped ones on the gate-centred OECTs, which required a preferential direction for the parylene C peeling as well as a fine tuning of the baking and peeling conditions to have an optimised yield.

The patches of PEDOT:PSS overlap the whole top area of the source and drain electrodes on a width of at least 10 $\mu$m in order to ensure the maximal injection of current in the channel. Based on a well-accepted mechanism [14], the channel conductance can be modulated upon source-and-drain/gate polarisation, such as the PEDOT semiconductor is dedoped from the PSS by reduction, supported by cations which penetrate the ionomer and chelate the sulfonates (see figure 2 a and b). Both the presence of labile cations and the device polarisation are essential but not sufficient in order to dedope of the PEDOT.

Indeed, first tests on our OECTs in hydrazine-free aqueous electrolytes of KCl showed weak conductance modulations in the studied voltage windows, even for concentrations up to 10 M of KCl (see figure 2 c). As published, the OECT dedoping efficiency depends on multiple parameters other than ionic concentration and applied voltages, such as the ratio of the channel area ($A_{ch}$) over the area of the gate ($A_g$) [15,16], the nature of the gate metal [17] and also the volume of PEDOT:PSS to dedope [18,19]. In order to achieve a significant dedoping of the whole PEDOT:PSS material with a small gate, the volume of PEDOT:PSS was kept as low as possible. Layers thinner than 20 nm of PEDOT:PSS were realised by spinning the formulation at velocities higher than 4000 rpm, followed by a subsequent spinning of deionised (DI) water to thinner it further before substrate baking. Despite the small thickness of polymer, rather good performances were obtained, reaching currents up to 60 $\mu$A. Also, the lack of hysteresis and the linearity of the curves indicate respectively a rapid dedoping at a scan rate of 5 mV/s and the currents to be ohmic (see figure 2 c).

Despite the low volume of PEDOT:PSS, the usage of Pt as gate metal and the small $A_{ch}/A_g$ poorly modulated the output-current density. It has been proposed that the electrochemical inertia of the gate metal in addition of its small size is at the origin of the lack of field-effect, causing the voltage to drop mainly at the platinum gate interface, screening the channel from polarisation [17] (see figure 2 d, upper skech). And in order to counterbalance this effect for the same $A_{ch}/A_g$, strategies such as using a fast redox-system gate like Ag/AgCl [17] or introducing a redox active analyte such as $H_2O_2$ in the electrolyte [15] have been used. In the case of $H_2O_2$ introduction, it has been reported that the field-effect on the OECT depends on the concentration of $H_2O_2$ introduced in the electrolyte, such as the OECT can be used as a $H_2O_2$ sensor down to $10^{-5}$ M (level of the mg/L) [15].

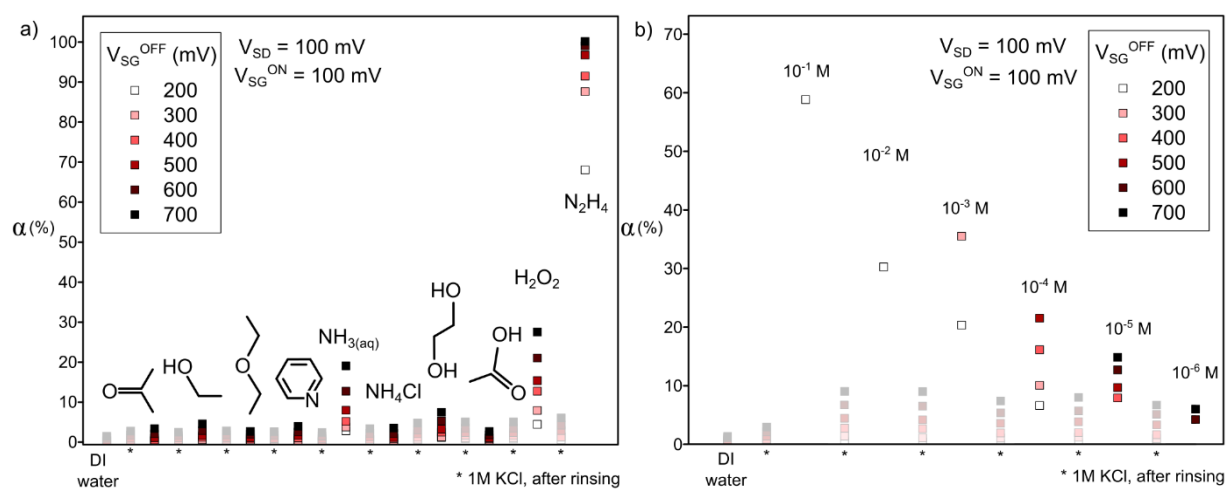

**Figure 3.** Selectivity and sensitivity of an OECT (conditions: source-centred, L = 5 $\mu$m and $R_{int}$ = 20 $\mu$m) to hydrazine at different source-gate potentials. a) Study of different 0.1 M concentrated analytes, diluted in 0.1 M KCl solutions. b) Study of $N_2H_4$, diluted in 1 M KCl solutions. Data on each graph are displayed from left to right in the chronological order.

In this study, we demonstrate the OECT to be far more selective to hydrazine. As a strong reductant, its electro-catalysed oxidation at the Pt gate supports the reduction of the PEDOT at the channel upon polarisation, creating $N_2$ and $N_2H_5^+$ as byproducts [20]. Considering the oxidation products which do not deposit on the gate, the chemical potential at the Pt-gate shall remain stable over time and the mirrored potential shall not depend on the sensor usage.

The electrical measurements of a source-centred OECT were performed with $N_2H_4$ at different electrical and chemical conditions, after the device being immersed overnight in deionised water (see figures 3).

If $\alpha(V_{SG}^{OFF}, V_{SG}^{ON}, V_{SD})$ is the amplitude of current modulation between the off-state and the on-state, defined such as:

$$\alpha = \frac{I(V_{SG}^{ON}; V_{SD}) - I(V_{SG}^{OFF}; V_{SD})}{I(V_{SG}^{ON}; V_{SD})} = 1 - \frac{I_{OFF}}{I_{ON}} \qquad (1)$$

and fixing the values $V_{SD}$ and $V_{SG}^{ON}$ at 100 mV, the sensor shows current density modulations $\alpha(V_{SG}^{OFF})$ below 3% with deionised water and 1 M concentrated KCl (see figure 3). This demonstrates that in the absence of an electroactive analyte, the output is barely sensitive to the concentration of KCl. Successively, different analytes have been introduced and tested on the OECT, followed by a rinsing and test on a 1 M KCl concentrated blank (see figure 3.a). After each analyte test, the sensor recovers down to a $\alpha = 5\pm2$ at $V_{SG}^{OFF} = 700$ mV, indicating the OECT to not be damaged by any analyte. The sensor displays a $\alpha$ for acetone, ethanol, diethylether, pyridine, ammonium chloride and acetic acid very near the one of the blanks, showing the insensitivity of the OECT for 0.1 M of these analytes in water. A slight modulation with 0.1 M of ethylene glycol was observed but without significant change in the PEDOT:PSS current density (since this analyte has the property to boost the conductivity of PEDOT:PSS [21], the lack of sensitivity to this molecule is of the highest importance). Moderated modulations with 0.1 M of aqua ammonia or hydrogen peroxide were observed. Confirming the applicability of $H_2O_2$ sensing with an OECT, it is worth noticing that other metabolites such as ammonia can also be detected at the same level. While 0.1 M of ammonia or hydrogen peroxide modulate with $\alpha = 25\pm5\%$ at $V_{SG}^{OFF} = 700$ mV, hydrazine promotes $\alpha = 65\pm5\%$ at as few as $V_{SG}^{OFF} = 200$ mV at the same concentration. The channel is completely closed with $I_{OFF} = 3\pm2$ nA when $V_{SG}^{OFF} = 700$ mV, resulting to $\alpha \approx 100\%$.

Dry hydrazine has the property to reduce p-doped polythiophenes such as P3HT [22]. An eventual thermodynamically favoured reduction of the PEDOT by $N_2H_4$ could explained the observed decrease of $I_{off}$ with the concentration of $N_2H_4$ (see figures 2 a). Nevertheless, this single property do not sufficiently justify the $V_{SG}^{OFF}$ dependency of the PEDOT conductivity, since the electrode polarisation under device operation promotes the electrolyte oxidation at the gate electrode and not at the PEDOT channel. The OECT sensor presents also an acute sensitivity to hydrazine (see figure 3.b). Successive dilutions of fresh hydrazine aqueous KCl solutions shows the sensor to be sensitive to at least $10^{-5}$ M of $N_2H_4$ at the studied voltages: The sensor responds positively for an applied $V_{SG}$ from 200 mV to 700 mV when exposed to an electrolyte of $N_2H_4$ of concentrations from 0.1 M to $10^{-5}$ M. The response $\alpha$ remains at the base level when immersed to a $10^{-6}$ M concentrated $N_2H_4$ electrolyte. We note that the sensor regeneration to $\alpha = 9\pm1$ is strongly limiting the device sensitivity at $10^{-6}$ M. Another operation protocol on the device rinsing step or on the electrical inputs could be thought to improve the sensor regeneration and therefore to increase the sensitivity range of the sensor, but it has not been explored in this work.

## 4. Discussion on the OECT structure-property relationship

Studies on millimetre-scale OECTs demonstrated that the sensitivity $\alpha$ of the devices to ions was highly dependent to gate electrode area [15]. Therefore, a parameter $\gamma = A_{ch}/A_g$ was defined as a preferred geometry parameter for the sensitivity of OECTs to ions at a given concentration and voltage [15]. Although this study enabled to correlate the device geometry to a dedoping strictly occurring in the channel, most practical micrometric bottom-contacts transistors requires a large area for the source-and-drain contacts to be exposed to the electrolyte [7]. This implies that the contact area has to be taken into account as counter-electrode for the gate, and therefore, impacts on α by both dedoping of the channel and the contacts, affecting both the channel conductivity and the Schottky barriers at the contacts. Since the conductivity and the Schottky barrier have a different dependency on the hole carrier density in the PEDOT [23], we believe that α of the OECTs can be severely impacted by parameters which are not exclusively channel-related, such as the contacts geometry.

To characterise it, tests of $10^{-2}$ M $N_2H_4$ in $10^{-1}$ M $KCl_{aq}$ were realised for all the device geometries, with $V_{SG}^{ON}$ = 700 mV and $V_{SG}^{OFF}$ = $V_{SD}$ = 100 mV (see figure 4). All the devices show a sensitivity α from 30 to 85% for both the source-centred geometries and the gate-centred ones (see figure 4 a). When referenced to 1 M $KCl_{aq}$, one sees the base modulation α without hydrazine to be located between 0 and 20% for both the source-centred and the gate-centred OECTs.

When plotting the sensitivity α against the geometric γ parameter of each device (see figure 4 b), one can distinguish a different amplitude for α of each OECT between operation with hydrazine and without hydrazine, showing all the devices to be operational and sensitive to it. Additionally, α of the OECT with reversed SD-polarisation are similar to the one of the direct SD-polarisation, indicating the insensitivity of α to the surface area of the hole injecting electrode, even for the largest $R_{int}$ source-centred devices. For the source-centred, a trend for α to decrease linearly with γ can be seen for a given channel length, and the slope of this trend seems to be channel length dependent.

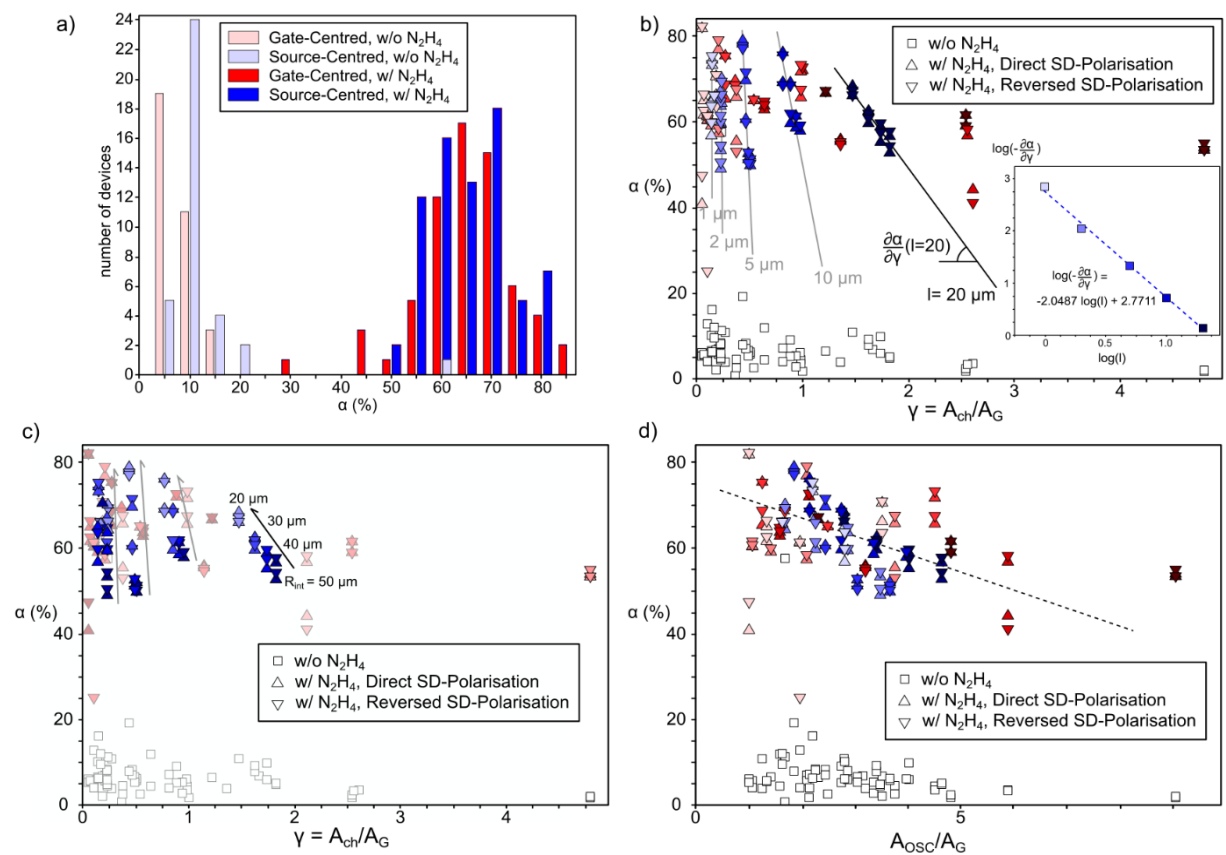

**Figure 4.** Study of the geometry-dependent sensitivity of OECTs. a) Statistical study on the sensitivity α for all geometries of OECTs, without and with $N_2H_4$. b) Plot of α(γ) showing the channel length dependency of the derivative for the source-centred OECTs (the inset plot shows the quadratic dependency). c) Plot of α(γ) showing the dependency of α with the radius of the central electrode for the source-centred OECTs. d) Plot of α($A_{OSC}/A_G$) showing a better correlation when considering a contact-area depending parameter (the dashed line is a guide for the eye). For all graphs, red accounts for gate-centred OECTs and blue for source-centred ones. In b) and d), the colour gradients are such as the lighter, the shorter channel and for c) the lighter, the smaller central electrode.

When plotting the logarithm of the slope the |α|(γ) curves for the different source-centred channel lengths, a slope equals to -2 can be seen (see the inset graph of figure 4 b), indicating the source-centred OECTs to have a sensitivity particularly depending on the channel length, such as the variation of α with γ increases in absolute value with $1/L^2$. But the fact that the 5 series of different channel length with different γ have comparable values of α under the same electrochemical conditions indicates that α depends on a geometric constraints other than $γ=A_{ch}/A_G$. When looking at the radius of the central electrode for each channel-length series of source-centred OECTs (see figure 4 c), one sees a monotonic trend for the sensitivity α to increase when the radius of the central electrode decreases. This indicates α to be source-and-drain-contact-area dependent and therefore the overlapping of the PEDOT:PSS with the contacts shall be considered for the design of electrochemical OECT sensors.

As an example, the figure 4 d shows that when we include the contact area in the discriminating geometric parameter, we see the series for both gate-centred and source-centred OECTs have the same sensitivity trends, with an overall trend for α to decrease with the area ratio PEDOT:PSS/gate ($A_{OSC}/A_G$).

In the end, α for the gate-centred series remain more dispersed around the trend than the source-centred ones, when plotted either against γ or $A_{OSC}/A_G$. This dispersion of α values for the gate-centred devices do not allow us to interpolate obvious trends with the channel lengths or the contact area as for the source-centred devices. Since the dimensions of both geometries are very comparable the one with the other (and so their γ and $A_{OSC}/A_G$ values), we believe the "C"-shaped PEDOT:PSS patches on the gate-centred devices are more subjects to variabilities than the round ones on the source-centred.

Regarding the trend for α to increase with the decrease of the contact area for source-centred OECTs, we would also recommend to diminish it to further increase the OECT sensitivity.

## 5. Conclusion

We report both a technological and a scientific results in the field of PEDOT:PSS-based organic electrochemical transistors for sensing. First, we demonstrate that this platform can selectively sense hydrazine, and as such, can be a promising structure to sense toxics in liquids. Second, a structure-property relationship showed the OECT to be highly contact sensitive. Future investigations on models taking into account the contact area of the devices in their functioning could help us to further understand their mechanism. This whole study was performed on micrometric concentric electrode OECTs: a set of engineered devices which might promote the development of higher definition electrochemical transistors.

## 6. Acknowledgments

The authors wish to thank the European Commission under the H2020-FETOPEN-2014-2015-RIA program, RECORD-IT project GA 664786, and the French National Nanofabrication Network RENATECH for financial supports. We thank also the IEMN cleanroom staff of their advices and support.